\documentclass[conference]{IEEEtran}
\IEEEoverridecommandlockouts
\usepackage{cite}
\usepackage{amsmath,amssymb,amsfonts}
\usepackage{algorithmic}
\usepackage[hidelinks]{hyperref}
\usepackage{graphicx}
\usepackage{textcomp}
\usepackage{xcolor}
\usepackage{paralist}
\usepackage{multirow}
\usepackage{acronym}
\usepackage[english]{babel}
\usepackage{soul,color}
\usepackage{caption}
\usepackage{subcaption}
\usepackage{footmisc}
\usepackage{hyperref}
\usepackage[export]{adjustbox}

\def\BibTeX{{\rm B\kern-.05em{\sc i\kern-.025em b}\kern-.08em
    T\kern-.1667em\lower.7ex\hbox{E}\kern-.125emX}}
\begin{document}

\title{Detecting Hardware Trojans in Microprocessors via Hardware Error Correction Code-based Modules}

\author{
\IEEEauthorblockN{Alessandro Palumbo}
\IEEEauthorblockA{\textit{CentraleSupélec, Inria, CNRS, IRISA, France} \\
alessandro.palumbo@inria.fr}
\and
\IEEEauthorblockN{Ruben Salvador}
\IEEEauthorblockA{\textit{CentraleSupélec, Inria, CNRS, IRISA, France} \\
ruben.salvador@inria.fr}
}

\maketitle

\IEEEpubidadjcol

\begin{abstract}
Software-exploitable Hardware Trojans (HTs) enable attackers to execute unauthorized software or gain illicit access to privileged operations. This manuscript introduces a hardware-based methodology for detecting runtime HT activations using Error Correction Codes (ECCs) on a RISC-V microprocessor. Specifically, it focuses on HTs that inject malicious instructions, disrupting the normal execution flow by triggering unauthorized programs. To counter this threat, the manuscript introduces a Hardware Security Checker (HSC) leveraging Hamming Single Error Correction (HSEC) architectures for effective HT detection. Experimental results demonstrate that the proposed solution achieves a 100\% detection rate for potential HT activations, with no false positives or undetected attacks. The implementation incurs minimal overhead, requiring only 72 \#LUTs, 24 \#FFs, and 0.5 \#BRAM while maintaining the microprocessor's original operating frequency and introducing no additional time delay.
\end{abstract}


\section{Introduction and related work}
The ongoing drive to lower production costs and shorter time-to-market has led to the globalization of design and manufacturing processes for modern integrated circuits (ICs)~\cite{DIGITIMES}. Design tasks for components and subsystems are often outsourced, third-party intellectual property cores (3PIPs) are commonly purchased, and the final chips are fabricated by external foundries~\cite{RKRK2013}. While this approach has effectively reduced both design time and cost, it has also led to a significant reduction in trust regarding the integrity of the produced ICs~\cite{TW2012}.

Ensuring trust among all participants in a globalized supply chain has become an increasingly complex challenge that is now virtually unattainable. As a result, various threats may emerge, including the overproduction or counterfeiting of integrated circuits (ICs), the improper use of licenses, and the introduction of Software-Exploitable Hardware Trojans (HTs)~\cite{space, towards}.

HTs are subtle, hard-to-detect alterations in a system designed to remain hidden for most of the time, only activating under certain (often rare) conditions to disrupt system behaviour by, e.g., performing attacks by accessing protected memory locations or stealing sensitive data~\cite{TK2010, xue2020ten}. Malicious actors can introduce such modifications, whether they are third-party IP providers, employees, CAD tools, mask suppliers, or silicon foundries.

Initially, HTs were considered more of an academic problem due to the challenges associated with their real-world implementation, which limited their potential use by attackers. However, recent research has demonstrated that they can be embedded in commercial microprocessors, enabling attackers to execute their own malicious software, modify running code, or gain root privileges~\cite{JM2012, TM2014, wang2012software}. For instance, a few years ago, the \textit{Rosenbridge} backdoor was discovered in a Via Technologies C3 processor~\cite{domas2018hardware}. Attackers could activate the backdoor through a specific instruction sequence, granting access to supervisor mode.

Several techniques have been developed for detecting HTs during the design phase, focusing on analyzing the system at the circuit level before deployment. These methods include logic testing~\cite{CYZ2017}, program run code obfuscation~\cite{deton, hdeton}, formal property verification~\cite{ZY2015}, side-channel analysis~\cite{LZ2017}, structural and behavioral analysis~\cite{ST2013, ST2012}, and machine learning~\cite{PALUMBO2022102543, ribes2024machine}. However, due to the inherently stealthy nature of HTs, detecting them before deployment is extremely challenging. This has led to the emergence of the \textit{Design-for-Trust} paradigm, which emphasizes \textit{system-level} techniques to construct trusted systems from potentially untrusted components~\cite{Sisejkovic, 7833075}. Additionally, HTs can be countered by enabling trusted software execution on systems that include untrusted microprocessor-based components~\cite{6569378, 5224959}.

This paper proposes a system-level solution for detecting HT activations at runtime in microprocessor-based systems, which could compel the microprocessor to execute malicious code. The presented approach integrates a Hardware Security Checker (HSC) composed of two Hardware Security Modules (HSMs) between the microprocessor and the main memory to monitor fetching activity. The HSC is \textit{configured} during the program installation in memory, utilizing information about the program's instructions and memory locations.

During the program execution, at runtime, the HSC verifies that the correct instructions are fetched from the correct memory addresses. This mechanism allows the HSC to detect HT activation at runtime, potentially compromising the microprocessor, the memory, or the bus. The proposed solution operates transparently without interrupting or interfering with the normal execution of the microprocessor. The HSC functions passively, ensuring the integrity of the protected program while remaining non-intrusive.

The HSC has been tested on a case study system based on a RISC-V microprocessor implemented on an FPGA, running a set of software benchmarks. The RISC-V ISA could implement the Physical Memory Protection (PMP) mechanism to control memory access and protect against unauthorized instruction execution or data manipulation~\cite{cheang2022verifying, ng2022realization}. However, PMP is not immune to attacks that exploit and modify its configuration~\cite{cheang2022verifying, shepherd2021lira}.
The results showed that the proposed HSC can detect 100\% potential HT activations without false alarms or undetected attacks. We observed an area overhead of less than 1\% in terms of \#LUTs and \#FFs, with 0.5 \#BRAM required and no reduction in operating frequency.

Some system-level design-for-trust methodologies are reported in~\cite{6569378, 5224959, 9080961, 9568291}. Some solutions~\cite{6569378, 5224959} assume the microprocessor is untrusted while the memory is trusted. In~\cite{6569378}, the checker monitors the validity of opcodes and control signals and the number of clock cycles required to execute an instruction. Meanwhile, the solution in~\cite{5224959} tracks the microprocessor's liveness and its privilege mode. However, neither of these solutions is designed to detect HTs that force the CPU to execute legitimate instructions without altering the privilege mode. Specifically, they do not detect scenarios where the microprocessor is compelled to execute an unintended program or access unauthorized memory locations, which our solution addresses.

Similar works~\cite{9080961, 9568291, palumbo2023improving} to this paper propose checkers to detect HT activations affecting the microprocessor, bus, and memory. The solution in~\cite{9080961} incurs significant area overhead and still experiences a non-zero false negative rate. The solution in~\cite{9568291} improves the overhead but leaves vulnerabilities when the accessed address is correct, yet the fetched instruction (opcode or operands) has been maliciously altered. In~\cite{palumbo2023improving}, the authors improved the HT detections by putting in parallel to the HSC a Hamming Single Error Correction (HSEC) circuitry. In~\cite{cms}, the authors presented a hardware-based methodology that monitors the microprocessors fetched instruction to detect Side-Channel Attacks (SCA). The approach presented in this manuscript integrates insights from these works by monitoring addresses and fetched instructions while adopting and evaluating HSEC codes to detect HT activations. This is achieved while maintaining minimal area and power overhead and without reducing the operating frequency.

This paper is organized as follows: Section~\ref{sec_model} presents the HT models targeted by our proposal; Section~\ref{sec_proposal} presents the proposed HT detection methodology and the details of the HSMs on which it relies; Section~\ref{sec_results} discusses the results of the proposed HSC, while Section~\ref{sec_secanalysis} discusses the security analysis; Section~\ref{sec_concl} concludes the paper.

\section{The considered threat models}
\label{sec_model}

This manuscript focuses on attacks that compel the CPU to execute unauthorized code. An HT could be embedded in the microprocessor, forcing the fetch unit to access instruction memory locations where malicious code is stored. Similarly, HTs might compromise the instruction memory or system bus, altering the referenced memory locations and thus facilitating the launch of a malicious program.

According to the classical HTs classification~\cite{TK2010}, no assumptions have been made about the triggering mechanism of the referred HTs. The proposed approach considers both triggered and always-on HTs. The presented methodology assumes that these HTs are introduced by a malicious Intellectual Property (IP) provider while the design team and foundry involved in the microprocessor's development are trusted. Therefore, the HSC is assumed to be reliable. Given that the attacker is the IP provider, we assume they have full knowledge of the hardware platform when inserting the HT.

The possible attack scenarios addressed by this proposal include:

\begin{enumerate}
  \item An HT in the bus or main memory forcing the fetching from a particular memory address, which in turn would contain an instruction forced by the attacker; 
  \item An HT in the bus or main memory that forces the fetching of an instruction from the legitimate program but at a different point in time than intended at design time.
\end{enumerate}

\section{The Proposed Security Solution}
\label{sec_proposal}

\begin{figure}[h]
\includegraphics[width=\linewidth]{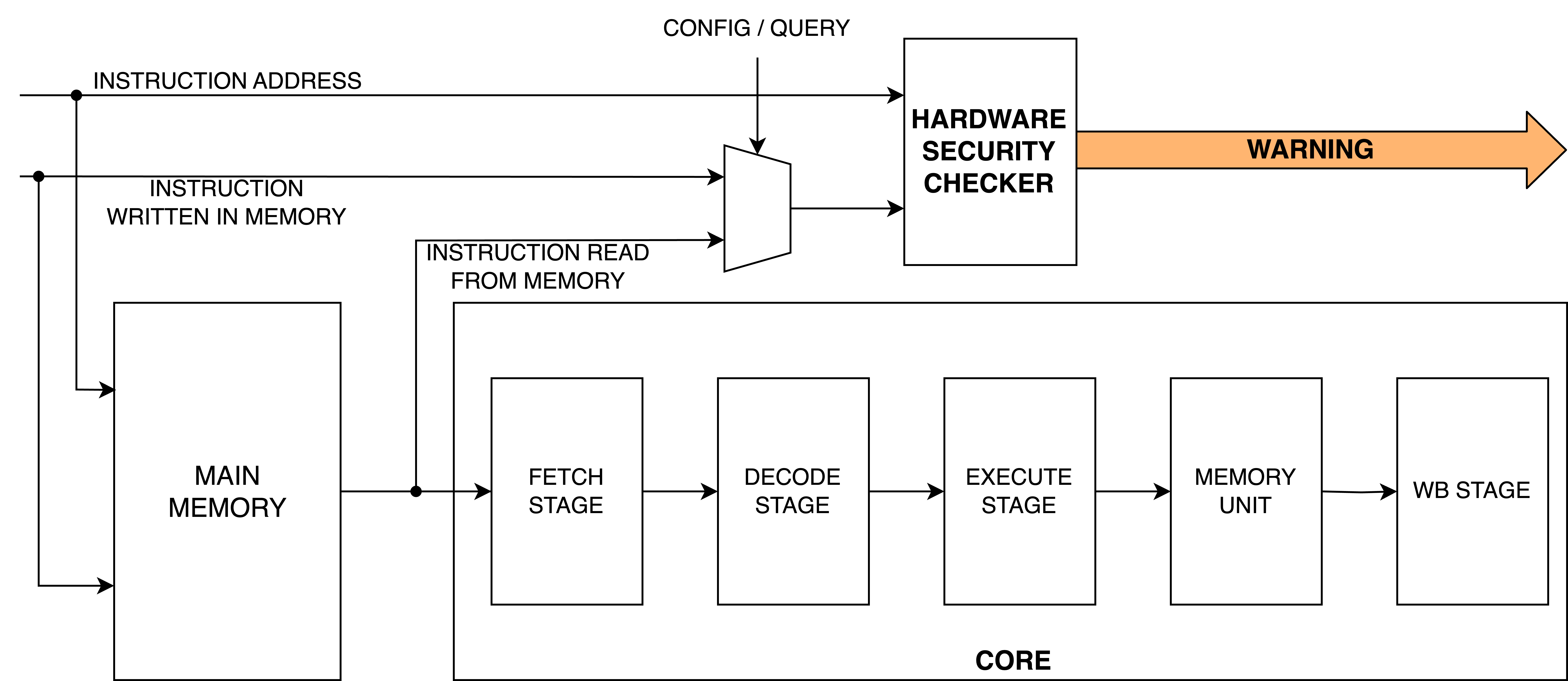}
\caption{The proposed protection architecture}
\label{fig:1}
\end{figure}

The proposed approach involves introducing an HSC between the microprocessor and the instruction memory (as illustrated in Figure~\ref{fig:1}) to detect HTs attempting to execute malicious code. The HSC is configured during the installation phase of the program(s) the system will execute. At runtime, the HSC monitors the instruction fetch phase to detect and signal any HT activation\footnote{The handling of the warning, such as via a non-maskable interrupt by the operating system, is outside the scope of this paper.}.

When the program is being installed into the main memory, the HSC operates in \textit{configuration} mode.
During this phase, it is programmed with the parity bits from the program instructions and the corresponding memory addresses. Once the program is installed, the HSC switches to \textit{query} mode. During this runtime phase, in parallel to each instruction fetch, the HSC checks if the accessed memory address and the fetched instruction parity bits match the previously configured data. Specifically, the HSC verifies that the accessed address belongs to the memory space of the running program and that the fetched instruction parity bits correspond to what was originally stored at that address during the installation phase.

\subsection{The Architecture of the Hardware Security Modules}

\begin{figure}[t]
\centering
\includegraphics[width=\linewidth]{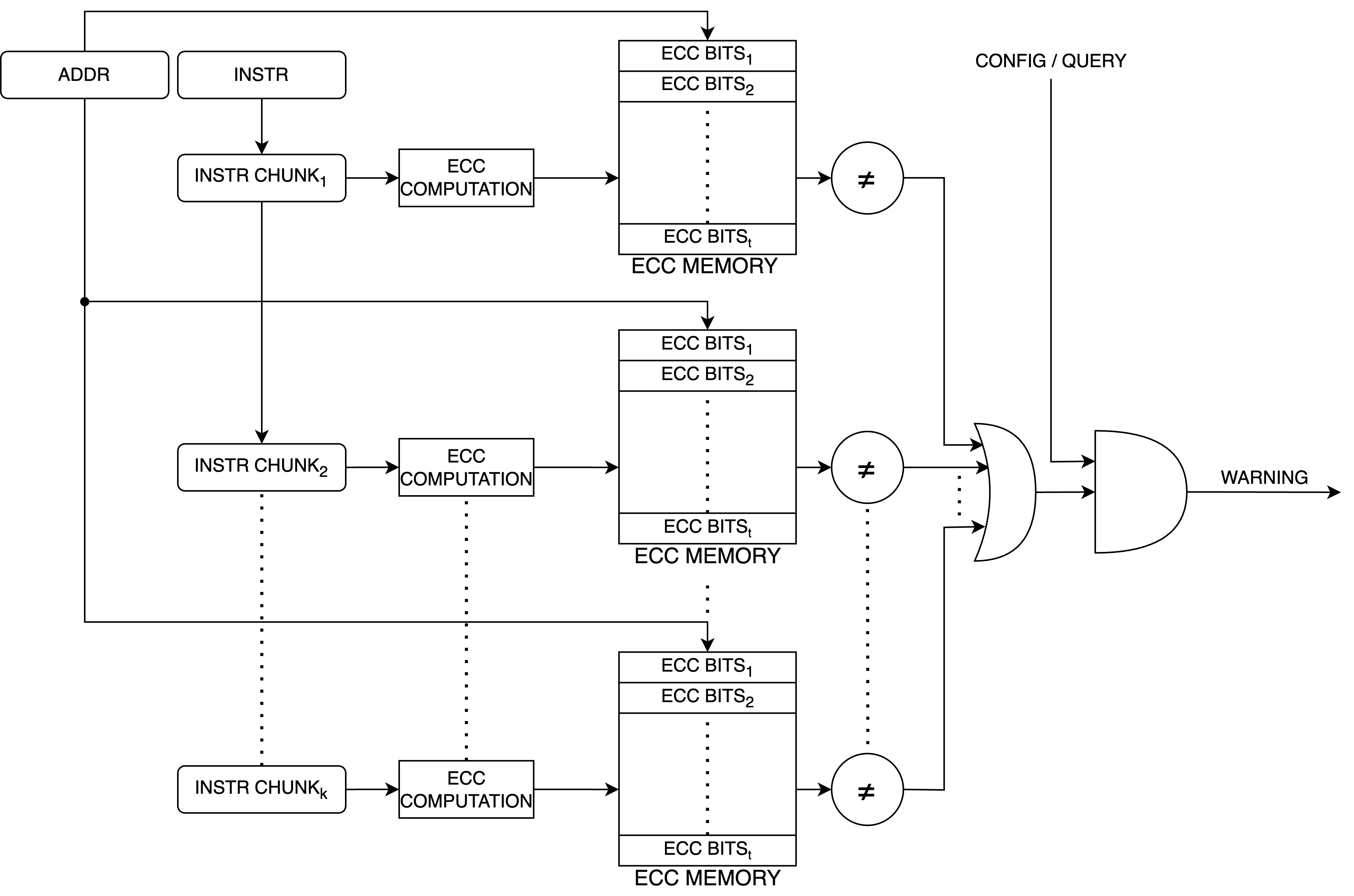}
\caption{The methodology of the proposed Hardware Security Module}
\label{fig:2}
\end{figure}

\begin{figure*}[t]
     \centering
     \begin{subfigure}[b]{0.49\textwidth}
         \centering
         \includegraphics[width=\linewidth]{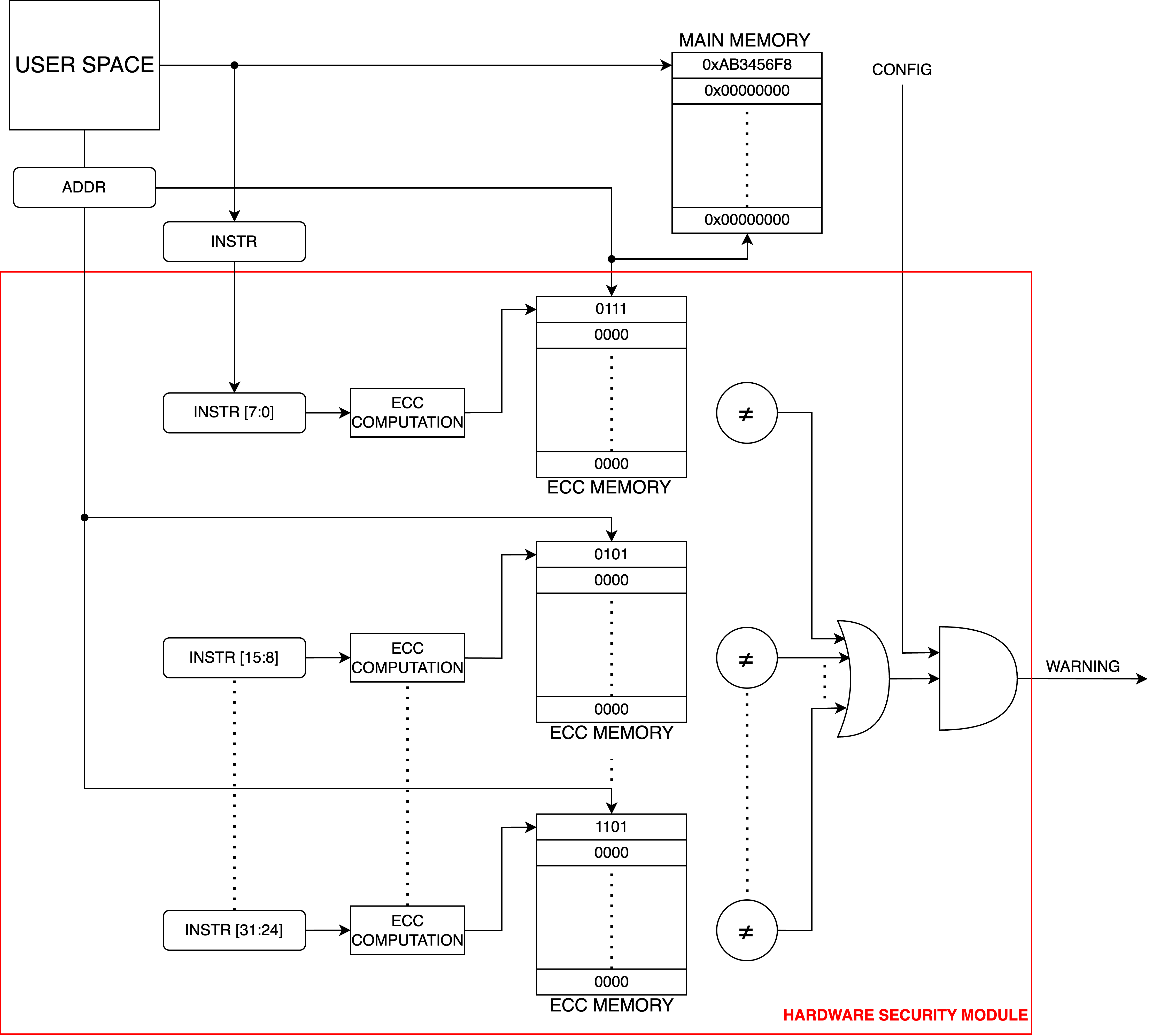}
          \caption{Writing the first program instruction}
         \label{fig:3.1}
     \end{subfigure}
     \hfill
     \begin{subfigure}[b]{0.49\textwidth}
         \centering
  \includegraphics[width=\linewidth]{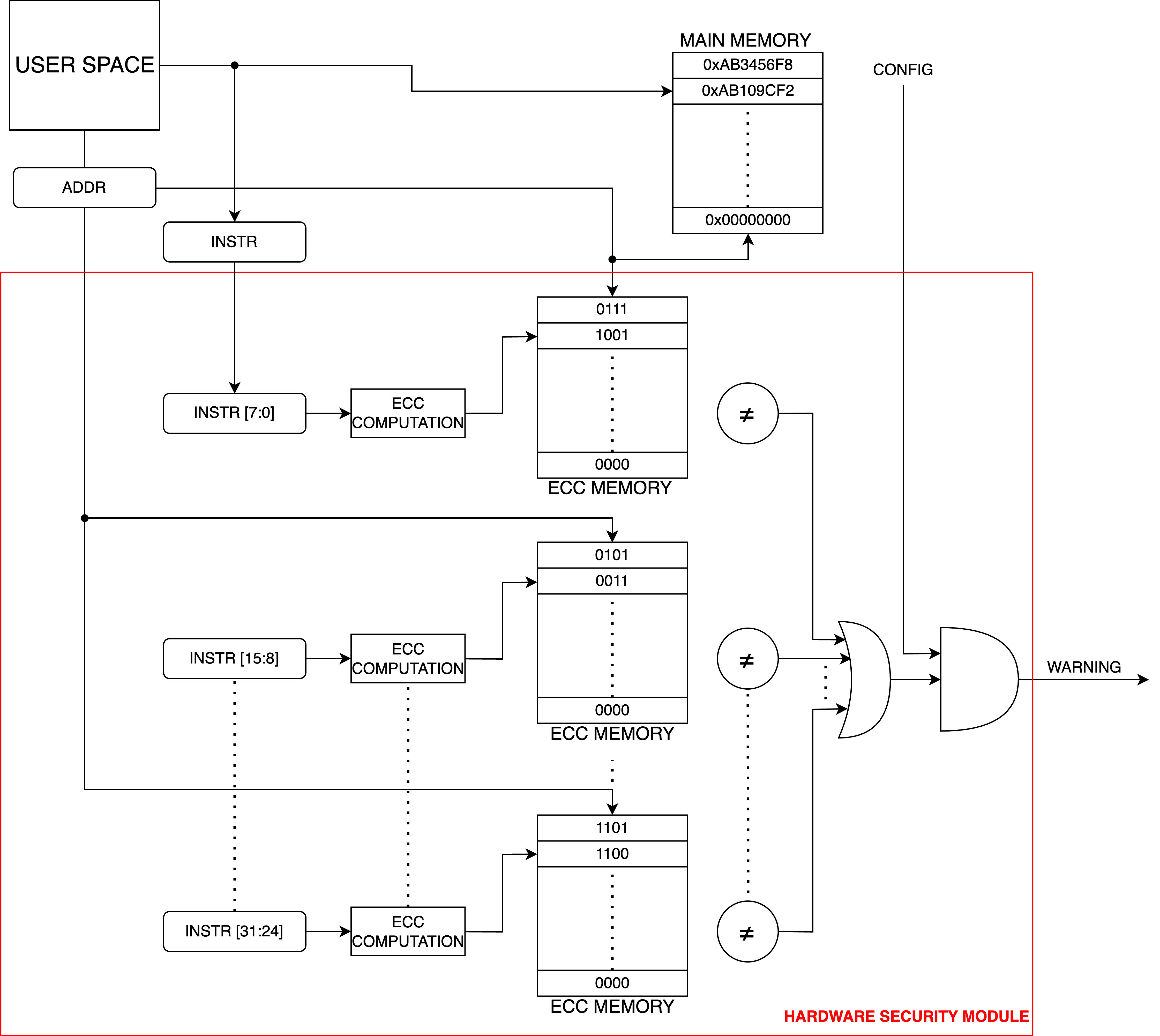}
 \caption{Writing the second program instruction}
  \label{fig:3.2}
     \end{subfigure}
       \caption{Configuration Methodology of the Hardware Security Module}
        \label{fig:3}
\end{figure*}

The architecture of the proposed methodology is illustrated in Figure~\ref{fig:2}: the HSMs implementing the HSC receive as input a memory address, an instruction, and the \texttt{CONFIGURE/QUERY} signal (which indicates whether the HSMs are operating in configure or query mode) and outputs a warning signal. In configure mode, the user provides both the addresses and instructions, meaning the program is being installed into the system's memory (Figure~\ref{fig:3.1}). In query mode, the addresses are the ones required by the microprocessor during the instruction fetch phase, while the instructions come from the main memory output (Figure~\ref{fig:3.2}).

The input addresses and instructions are divided into chunks (in a manner detailed in Subsection~\ref{sub_sec_config}) and used to point to a set of memories within the HSMs during both the configure and query modes. During the configuration phase, the ECC sequences (i.e., the checking bits) are calculated for each instruction chunk. These checking bits are stored in dedicated memories, where each instruction chunk has a corresponding address. During the query phase, the checking bits of the fetched instruction are recalculated and compared with those stored previously during the configuration phase at the memory address from which the microprocessor is fetching the current instruction. By incorporating these checking bits, the system can detect errors (instruction bit flips) and HT activation using the same hardware blocks, ensuring efficient and robust security checking.
In query mode, the HSMs raise a warning if the calculated checking bits do not match the stored ones.

\begin{table}[h!]
\centering
\caption{ECC Types and Parameters} \label{tab_ecc_details}
\begin{tabular}{l|c|c|c}
\textbf{ECC Type} & $\textbf{n/k}$ & $\textbf{k}$ & $\textbf{p}$ \\ \hline
\texttt{HSEC32} & 32 & 1 & 6 \\ \hline
\texttt{HSEC16} & 16 & 2 & 5  \\ \hline
\texttt{HSEC8} & 8 & 4 & 4  
\end{tabular}%
\end{table}

\subsection{Configuration and Usage of the Hardware Security Module}
\label{sub_sec_config}
The HSMs receive an address and an instruction as inputs. 
These two inputs are concatenated and then divided into a series of vector chunks (\texttt{INSTR CHUNK$_1$} to \texttt{INSTR CHUNK$_{k}$} in Figure~\ref{fig:2}). Specifically, the number of memories in the HSMs is $k$, the \textit{fragmentation factor}, while $n$ represents the bit-size of addresses and instructions in the architecture under consideration. \texttt{INSTR CHUNK$_1$} is defined as the chunk containing the first $n/k$ bits of the address coupled with the first $n/k$ bits of the instruction. Likewise, \texttt{INSTR CHUNK$_2$} includes the second $n/k$ bits of the address and the second $n/k$ bits of the instruction, and this pattern continues. These chunks are then inputted to ECC architectures, providing parity bits that will be stored in ECC's corresponding memories.

When the HSMs operate in configure mode, the user space provides both addresses and instructions while setting up the program. Figure~\ref{fig:3} illustrates an example of the configuration process for the program’s two initial instructions within a system with 32-bit addresses and instructions and an HSM with $n=32$ and $k=4$. It is important to note that different instructions may lead one or more vector chunks to reference the same parity bits. Consequently, different ECC memory locations can contain the same parity bits.

\begin{figure}
\centering
\includegraphics[width=\linewidth]{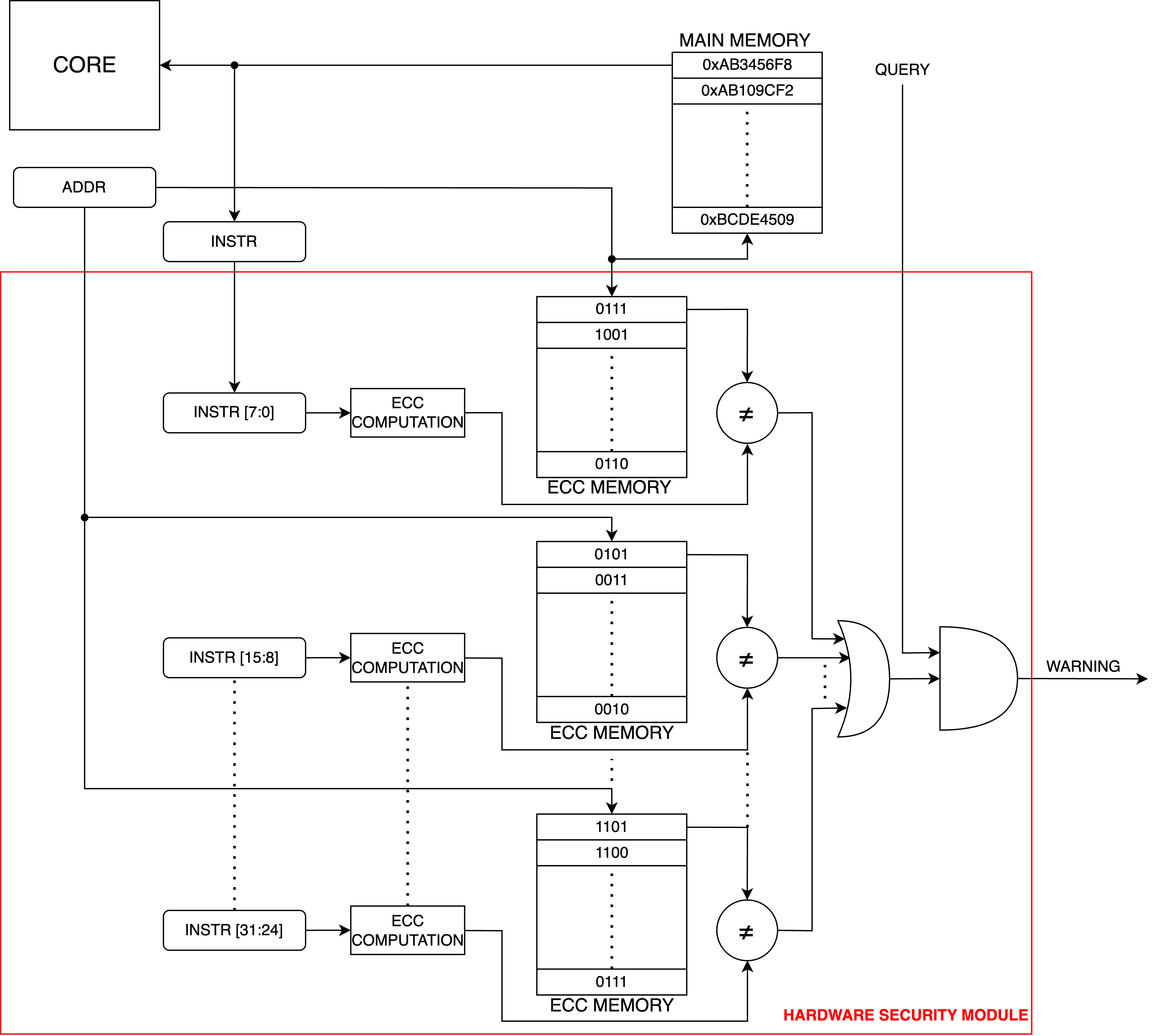}
\caption{Querying Methodology of the Hardware Security Module}
\label{fig:4}
\end{figure}

When the HSMs are in query mode, the address is provided by the microprocessor, and after retrieving it from the main memory, the corresponding instruction is also obtained. The $k$ vector chunks are generated in the same manner as previously described; however, in this phase, the contents of the ECC Memory are read (as opposed to being written in configure mode). The HSMs compute the parity bits for the instruction retrieved from the main memory and compare them with the ones previously stored in the ECC memory at the address from the microprocessor. If this comparison indicates a discrepancy, the HSM raises an alarm. Figure~\ref{fig:4} provides an example query process for the program’s first instruction.

\section{Experimental Results}
\label{sec_results}
This section describes the hardware platform and benchmark programs used, presents the results obtained from the experiments, and provides a detailed security analysis.
Three Hamming Single Error Correction (HSEC) functions were assessed regarding detection rates. Their architectural details are summarized in Table~\ref{tab_ecc_details}, where $n/k$ indicates the number of bits in each chunk, $k$ denotes the fragmentation factor (representing the number of chunks managed by the HSMs), and $p$ is the number of bits per ECC memory entry.

\subsection{Experimental setup}
\label{sub_sec_results}

\begin{table}[t]
\begin{center}
\caption{The considered benchmark programs}\label{benchmarks_ran}
\begin{tabular}{l|r}

{\textbf{Benchmarks}}           & {\textbf{\#32-bit Instruction (I)}} \\ \hline
\texttt{Coremark (CM)}              &1288           \\ 
\texttt{Matrix Multiplication (MM)} & 216                \\ 
\texttt{Quick Sort (QS)}            & 1023                  \\ 
\texttt{RSort (RS)}                 & 4466                \\ 
\texttt{Sha (SHA)}                  & 516               \\ 
\end{tabular}
\end{center}
\end{table}

For benchmarking, the programs listed in Table~\ref{benchmarks_ran} were selected, along with their respective counts of assembly instructions. The experimental analysis utilized the RI5CY~\cite{7864441} core from the PULPINO architecture, a 32-bit ultra-low-power processing platform designed for Internet of Things applications~\cite{traber2016pulpino}. When implemented on a Xilinx Artix XC7A35T FPGA~\cite{RISCYFPGA}, the RI5CY core requires 15,314 LUTs and 9,881 FFs, achieving a clock frequency of approximately 50 MHz. Using the same board and core, Table~\ref{tab_resources} summarizes the resources utilized by the HSC and the percentage overheads associated with different ECC architectures. These results are compared with a previous HSC proposal~\cite{palumbo2023improving}, which employs ECCs to enhance the detection of HT activations.
The HTs were emulated by randomly modifying the instruction memory address from which the microprocessor fetches instructions at runtime. Specifically, HTs capable of hijacking the execution flow toward a new program were considered. These HTs force the selected memory address to fall outside the memory space of the program currently under execution.
A total of 10,000 randomly generated HT activations were run, and the false negative (FN) rate was calculated as the ratio of runs in which the checker failed to raise an alarm to the total number of runs. Similarly, another 10,000 runs were performed without activating any HTs, and the false positive (FP) rate was measured as the ratio of runs in which the HSMs raised an alarm to the total number of runs. FPs, corresponding to false alarms, are theoretically and mathematically impossible due to the inherent properties of the error detection mechanism. Specifically, the HSMs approach relies on deterministic parity checks that uniquely map to the expected instruction execution flow. Since the ECC mechanism is designed to detect only deviations from this predefined structure, it cannot erroneously flag a correctly executed instruction sequence as an anomaly. In other words, given that the error-checking codes are computed based on a fixed and predictable transformation of the instruction data, any modification required to produce a false positive would necessitate an alteration that coincidentally results in a valid parity match---an event with negligible probability. Therefore, under ideal conditions without external sources of hardware faults or transient bit flips unrelated to the attack model, FPs cannot occur. 
The results of this experiment are summarized in Tables~\ref{HSEC_HT_m1} and~\ref{HSEC_HT_m2}. 
Such tables seem to reveal an anomalous trend in the False Negative (FN) rates for the \texttt{HSEC16} implementation compared to \texttt{HSEC32} and \texttt{HSEC8}. Theoretically, it is expected that:
\texttt{HSEC32}, calculated on a single 32-bit block, would exhibit the highest FN rate because it evaluates a larger data chunk and provides a more robust parity check;
\texttt{HSEC16}, calculated on two separate 16-bit blocks, would show an intermediate FP rate;
\texttt{HSEC8}, calculated on four separate 8-bit blocks, would have the lowest FN rate due to finer granularity and an increased number of independent calculations.
However, the results indicate that the FN rate for \texttt{HSEC16} is higher than for \texttt{HSEC32} and \texttt{HSEC8}. Such a trend has been confirmed by similar tests running this time CRC algorithms on the same chunk divisions (\texttt{CRC32}, \texttt{CRC16}, \texttt{CRC8}), as shown in Tables~\ref{CRC_HT_m1} and~\ref{CRC_HT_m2}.

\begin{table}[t]
\begin{center}
\caption{FP and FN rates when the HT modifies the accessed instruction memory location (Threat model 1 reported in Section~\ref{sec_model})}\label{HSEC_HT_m1}
\begin{tabular}{l|rr|rr|rr}
\multirow{2}{*}{\textbf{Benchmarks}}  & \multicolumn{2}{c|}{\texttt{HSEC32}} & \multicolumn{2}{c|}{\texttt{HSEC16}} & \multicolumn{2}{c}{\texttt{HSEC8}}
\\
             & FP   & FN    & FP    & FN    & FP    & FN           \\ \hline
\texttt{CM}  & 0\%  & 1.66\%   & 0\%   & 3.20\%   & 0\%   & 0\%    \\
\texttt{MM}  & 0\%  & 1.52\%   & 0\%   & 3.28\%   & 0\%   & 0\%     \\
\texttt{QS}  & 0\%  & 1.40\%   & 0\%   & 3.15\%   & 0\%   & 0\%     \\
\texttt{RS}  & 0\%  & 1.55\%   & 0\%   & 3.00\%   & 0\%   & 0\%     \\
\texttt{SHA} & 0\%  & 1.61\%   & 0\%   & 2.97\%   & 0\%   & 0\%     \\
\hline
AVG          & 0\%  & 1.548\%   & 0\%   & 3.12\%   & 0\%   & 0\% 
\end{tabular}
\end{center}
\end{table}

\begin{table}[t]
\begin{center}
\caption{FP and FN rates when the HT modifies the fetched instruction (Threat model 2 reported in Section~\ref{sec_model})}\label{HSEC_HT_m2}
\begin{tabular}{l|rr|rr|rr}
\multirow{2}{*}{\textbf{Benchmarks}}  & \multicolumn{2}{c|}{\texttt{HSEC32}} & \multicolumn{2}{c|}{\texttt{HSEC16}} & \multicolumn{2}{c}{\texttt{HSEC8}} 
\\
             & FP   & FN    & FP    & FN    & FP    & FN         \\ \hline
\texttt{CM}  & 0\%  & 1.64\%   & 0\%   & 3.55\%   & 0\%   & 0.12\%    \\
\texttt{MM}  & 0\%  & 1.72\%   & 0\%   & 2.52\%   & 0\%   & 0.51\%      \\
\texttt{QS}  & 0\%  & 1.61\%   & 0\%   & 2.93\%   & 0\%   & 0.07\%      \\
\texttt{RS}  & 0\%  & 1.60\%   & 0\%   & 3.12\%   & 0\%   & 0.02\%      \\
\texttt{SHA} & 0\%  & 1.71\%   & 0\%   & 3.21\%   & 0\%   & 0.14\%    \\
\hline
AVG          & 0\%  & 1.656\%   & 0\%   & 3.066\%   & 0\%   & 0.172\% 
\end{tabular}
\end{center}
\end{table}

\begin{table}[t]
\begin{center}
\caption{FP and FN rates when the HT modifies the accessed instruction memory location (Threat model 1 reported in Section~\ref{sec_model})}\label{CRC_HT_m2}
\begin{tabular}{l|rr|rr|rr}
\multirow{2}{*}{\textbf{Benchmarks}}  & \multicolumn{2}{c|}{\texttt{CRC32}} & \multicolumn{2}{c|}{\texttt{CRC16}} & \multicolumn{2}{c}{\texttt{CRC8}}
\\
             & FP   & FN    & FP    & FN    & FP    & FN         \\ \hline
\texttt{CM}  & 0\%  & 0.09\%   & 0\%   & 0.83\%   & 0\%   & 0.09\%    \\
\texttt{MM}  & 0\%  & 0.44\%   & 0\%   & 0.44\%   & 0\%   & 0.44\%      \\
\texttt{QS}  & 0\%  & 0.01\%   & 0\%   & 0.01\%   & 0\%   & 0.01\%      \\
\texttt{RS}  & 0\%  & 0.03\%   & 0\%   & 0.34\%   & 0\%   & 0.03\%      \\
\texttt{SHA} & 0\%  & 0.18\%   & 0\%   & 1.11\%   & 0\%   & 0.18\%    \\
\hline
AVG          & 0\%  & 0.015\%   & 0\%   & 0.546\%   & 0\%   & 0.15\% 
\end{tabular}
\end{center}
\end{table}

\begin{table}[t]
\begin{center}
\caption{FP and FN rates when the HT modifies the fetched instruction (Threat model 2 reported in Section~\ref{sec_model})}\label{CRC_HT_m1}
\begin{tabular}{l|rr|rr|rr}
\multirow{2}{*}{\textbf{Benchmarks}}  & \multicolumn{2}{c|}{\texttt{CRC32}} & \multicolumn{2}{c|}{\texttt{CRC16}} & \multicolumn{2}{c}{\texttt{CRC8}} 
\\
             & FP   & FN    & FP    & FN    & FP    & FN           \\ \hline
\texttt{CM}  & 0\%  & 1.66\%   & 0\%   & 3.20\%   & 0\%   & 0\%    \\
\texttt{MM}  & 0\%  & 1.52\%   & 0\%   & 3.28\%   & 0\%   & 0.01\%     \\
\texttt{QS}  & 0\%  & 1.40\%   & 0\%   & 3.15\%   & 0\%   & 0\%     \\
\texttt{RS}  & 0\%  & 1.55\%   & 0\%   & 3.00\%   & 0\%   & 0\%     \\
\texttt{SHA} & 0\%  & 1.61\%   & 0\%   & 2.97\%   & 0\%   & 0\%     \\
\hline
AVG          & 0\%  & 1.548\%   & 0\%   & 3.12\%   & 0\%   & 0.002\% 
\end{tabular}
\end{center}
\end{table}

In these tests, the FN rates followed the same trend, suggesting that the discrepancies are not specific to the ECC type (Hamming or CRC) but are instead related to intrinsic factors such as block segmentation and the statistical distribution of instructions. The \texttt{HSEC16} (and the \texttt{CRC16}) configuration relies on two 16-bit blocks for checking bit calculations. After these experiments, we can conclude that the data (i.e., \texttt{INSTR CHUNKS$_{1}$} and \texttt{INSTR CHUNKS$_{2}$}) in these blocks exhibit a high degree of correlation or overlap during the benchmark execution, reducing the effectiveness of checking bits, leading to a higher FN rate.

\begin{figure}[h!]
\centering
\includegraphics[width=\linewidth]{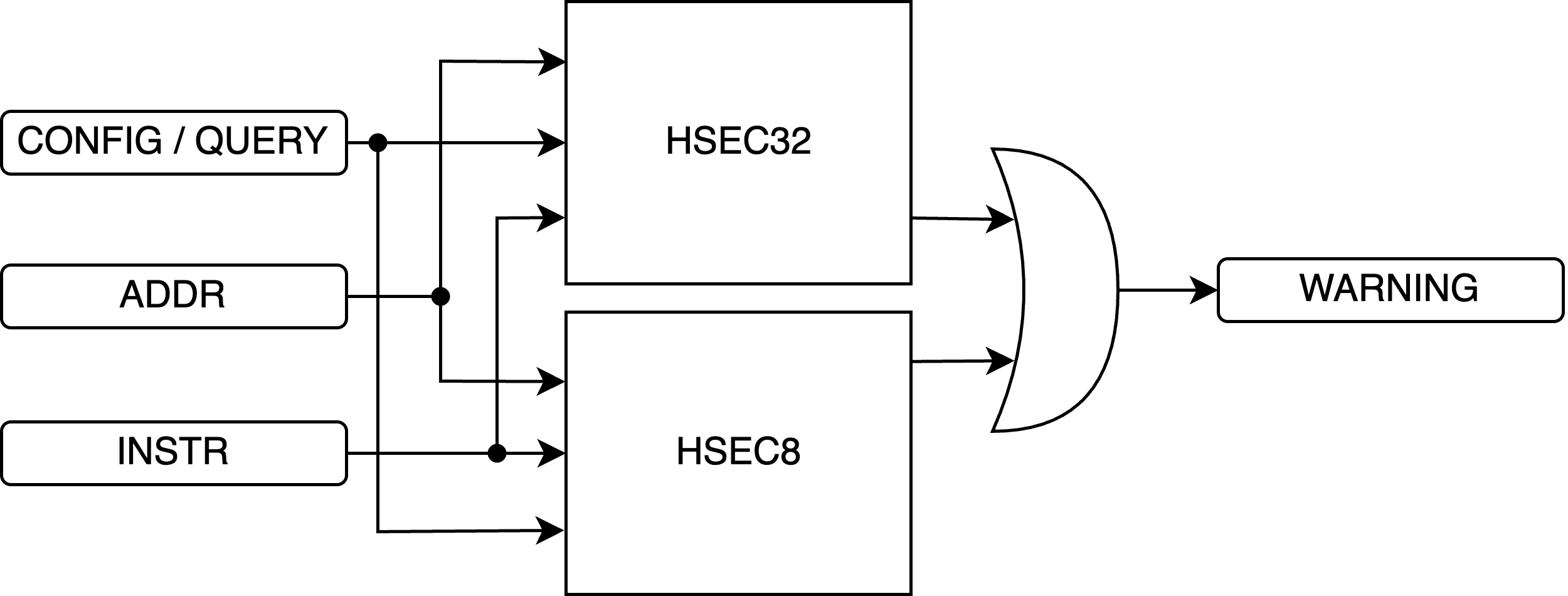}
\caption{The structure of the proposed HSC}
\label{fig_proposed_HSC}
\end{figure}

Given these findings, the proposed HSC combines \texttt{HSEC32}- and \texttt{HSEC8}-based HSMs to optimize detection accuracy while minimizing FN. The proposed HSC is illustrated in Figure~\ref{fig_proposed_HSC}.

\begin{table*}[h]
\begin{center}
\caption{Resource occupation and working frequency}
\label{tab_resources}
\begin{tabular}{l|rrr|r||rrr|r}
\multirow{2}{*}{\textbf{Benchmarks}} & \multicolumn{4}{c||}{\textbf{Proposed solution}} & \multicolumn{4}{c}{\textbf{Solution in~\cite{palumbo2023improving}}}
\\
                            & \#LUTs   & \#FFs    & \#BRAMs        & F. (MHz) & \#LUTs   & \#FFs    & \#BRAMs        & F. (MHz) \\ \hline
\texttt{CM}    & 72 (0.47\%)& 24 (0.24\%) & 0.5 & 560 MHz & 82 (0.53\%) & 31 (0.31\%) & 8.5 & 275 MHz \\
\texttt{MM}    & 72 (0.47\%)& 24 (0.24\%) & 0.5 & 560 MHz & 82 (0.53\%) & 31 (0.31\%) & 8.5 & 275 MHz  \\
\texttt{QS}    & 72 (0.47\%)& 24 (0.24\%) & 0.5 & 560 MHz & 82 (0.53\%) & 31 (0.31\%) & 8.5 & 275 MHz   \\
\texttt{RS}    & 72 (0.47\%)& 24 (0.24\%) & 0.5 & 560 MHz & 82 (0.53\%) & 31 (0.31\%) & 9.5 & 275 MHz\\
\texttt{SHA}   & 72 (0.47\%)& 24 (0.24\%) & 0.5 & 560 MHz & - & - & - & - 
\end{tabular}
\end{center}
\end{table*}

Finally, targeting an FPGA implementation, the overhead introduced by the proposed HSC in terms of used resources and working frequency has been evaluated. Table~\ref{tab_resources} reports the number of LUTs, FFs, and BRAM blocks required, as well as the maximum working frequencies, by the proposed HSC and compares them with previous work~\cite{palumbo2023improving}. The overhead in terms of additional FFs and LUTs is negligible between the two solutions. In addition, both of them do not slow down the microprocessor that works at 50MHz. On the other hand, our solution requires much less BRAMs. 

We also compare undetected and false alarm rates. Table~\ref{tab_last} reports the rate of the threat model 2 reported in Section~\ref{sec_model} (both the solutions do not have any FP or FN on the threat model 1 reported in Section~\ref{sec_model}). \textbf{One of the key contributions of this work is demonstrating that ECCs are capable of detecting and correcting errors and fully capable of detecting runtime security attacks, such as HT activations}. This is a significant advancement compared to previous work~\cite{palumbo2023improving}, which showed that combining ECCs with additional hash functions could enhance attack detection. Here, \textbf{we demonstrate that ECCs alone can effectively detect attacks}, eliminating the need for additional hardware complexity. This finding could have profound implications for microprocessor-based systems:
\begin{itemize}
    \item \textbf{Reuse of Existing ECC Blocks:} Modern microprocessors often include ECC-based error detection and correction mechanisms to ensure data integrity. Our results show that these existing ECC blocks can also be repurposed or extended to detect security threats, providing a dual-use capability without significant additional resource overhead.
    \item \textbf{Streamlined Design:} By relying solely on ECCs, our approach simplifies the hardware implementation, making it more cost-effective and easier to integrate into existing microprocessor architectures. This is especially valuable for low-power or resource-constrained environments like IoT devices or embedded systems.
\item \textbf{Comprehensive Security Coverage:} Unlike the previous solution, which required hash functions to complement ECCs for effective attack detection, we demonstrate that ECCs are sufficient to detect attacks with high accuracy. This fully exploits the potential of ECCs, turning them into a robust tool for runtime security monitoring.
\item \textbf{Scalability and Adaptability:} The ability to detect errors and attacks using the same ECC-based hardware ensures that the solution can be easily adapted to various microprocessor architectures without redesigning core security components.
\end{itemize}

\begin{table}[t]
\begin{center}
\caption{FP and FN rates (Threat model 2 reported in Section~\ref{sec_model})}\label{tab_last}
\begin{tabular}{l|rr|rr}
\multirow{2}{*}{\textbf{Benchmarks}}      & \multicolumn{2}{c}{\textbf{Proposed solution}} & \multicolumn{2}{c}{\textbf{Solution in~\cite{palumbo2023improving}}}  \\
                & FP          & FN       & FP         & FN           \\ \hline
\texttt{CM}     & 0\%      & 0.12\%   & 0\%      & 0.11\%  \\
\texttt{MM}     & 0\%      & 0.44\%   & 0\%      & 0.34\%       \\
\texttt{QS}     & 0\%      & 0.01\%   & 0\%      & 0.08\%      \\
\texttt{RS}     & 0\%      & 0.12\%   & 0\%      & 0.05\% \\
\texttt{SHA}    & 0\%      & 0.14\%   & -      & - \\
\hline
AVG             & 0\%      & 0.166\%      & 0\%     & 0.145\% 
\end{tabular}
\end{center}
\end{table}

\section{Security analysis}
\label{sec_secanalysis}

The presented experimental results demonstrate that the proposed HSC allows the detection of 100\% of the runtime activations of HTs that try to force the CPU to execute malicious programs installed in instruction memory locations outside the memory space of the running program as well as 100\% of HTs that try to force the CPU to execute a legit instruction but in a different moment respect to the designed one. Furthermore, the proposed HSC never incurs false alarms. 
It is worth mentioning that, as it has already been discussed, the effectiveness of the proposed solution is independent of the triggering mechanism of the HT, i.e., combinational/sequentially triggered, externally activated, time-bombs and always-on, and of the design stage during which the HT has been inserted. 

The proposed solution could be defeated by denial-of-service HTs that modify the execution flow of the legal program. We identified two possible scenarios: i) HTs that halt the system by maliciously making the CPU fetch always the same legal instruction (or sequence of legal instructions) from legal memory locations, and ii) HTs that halt the system by making it crash by fetching a legal instruction from a memory location belonging to the authorized program but at the wrong time or in the wrong order, e.g., fetching a jump instruction too early during the execution flow.

\section{Conclusion}
\label{sec_concl}

We presented a security architecture to protect microprocessor-based systems against HT attacks. We integrated our proposal within a system featuring a RISC-V processor implemented on an FPGA device and running a set of software benchmarks. Our proposal can detect 100\% of possible Hardware Trojan (HT) activations without false or undetected alarms. We measured a LUT and FF overhead of less than 1\%, with 0.5 \#BRAMs required and no working frequency reduction.

\bibliography{biblio}
\bibliographystyle{ieeetr}

\end{document}